\documentclass[aps,showpacs,twocolumn,groupedaddress,floatfix]{revtex4}
\usepackage{graphicx}  % needed for figures
\usepackage{dcolumn}   % needed for some tables
\usepackage{bm}        % for math
\usepackage{amssymb}   % for math
\usepackage{lineno}
\usepackage{slashed}

\begin{document}
%\MakeLineNo
% The following information is for internal review,
% please remove them for submission
%\leftline{To be submitted to PRL}

% the following line is for submission, including submission to the arXiv!!
\hspace{5.2in} \mbox{Fermilab-Pub-07/620-E}

\title{ Search for sneutrino particles in $e+\mu$ final states in $p\overline{p}$ collisions at $\sqrt{s} =1.96$ TeV }
%\author{D0 collaboration}
% LIST_OF_AUTHORS_R2.TEX               10/09/07(b)          
%
\author{V.M.~Abazov$^{36}$}
\author{B.~Abbott$^{76}$}
\author{M.~Abolins$^{66}$}
\author{B.S.~Acharya$^{29}$}
\author{M.~Adams$^{52}$}
\author{T.~Adams$^{50}$}
\author{E.~Aguilo$^{6}$}
\author{S.H.~Ahn$^{31}$}
\author{M.~Ahsan$^{60}$}
\author{G.D.~Alexeev$^{36}$}
% ...

\affiliation{\vspace{0.1 in}(The D\O\ Collaboration)\vspace{0.1 in}}
% ...
\affiliation{$^{6}$University of Alberta, Edmonton, Alberta, Canada,
                Simon Fraser University, Burnaby, British Columbia, Canada,
                York University, Toronto, Ontario, Canada, and
                McGill University, Montreal, Quebec, Canada}
% ...
\affiliation{$^{29}$Tata Institute of Fundamental Research, Mumbai, India}
% ...
\affiliation{$^{31}$Korea Detector Laboratory, Korea University, Seoul, Korea}
% ...
\affiliation{$^{36}$Joint Institute for Nuclear Research, Dubna, Russia}
% ...
\affiliation{$^{50}$Florida State University, Tallahassee, Florida 32306, USA}
% ...
\affiliation{$^{52}$University of Illinois at Chicago,
                Chicago, Illinois 60607, USA}
% ...
\affiliation{$^{60}$Kansas State University, Manhattan, Kansas 66506, USA}
% ...
\affiliation{$^{66}$Michigan State University,
                East Lansing, Michigan 48824, USA}
% ...
\affiliation{$^{76}$University of Oklahoma, Norman, Oklahoma 73019, USA}
% ...
  % input Dzero author list
%Send comments to d0-run2eb-012@fnal.gov \vfill by XXX XX, XXXX  \\
%Primary authors: X.~Bu, L.~Han, Y.~Liu, H.~Yin \vfill  Version: 17
\date{November 20, 2007}

\begin{abstract}

  We report a search for $R$-parity violating production and decay of 
   sneutrino particles in the 
  $e\mu$ final state with 1.04$\pm$0.06 fb$^{-1}$ of
   data collected with the D$0$ detector at the Fermilab
  Tevatron Collider in 2002--2006.
  Good agreement between the data and the standard model prediction is
  observed. With no evidence for new physics, we set limits
  on the $R$-parity violating couplings $\lambda^{\prime}_{311}$ and $\lambda_{312}$
  as a function of sneutrino mass.

\end{abstract}

\pacs{14.80.Ly, 12.60.Jv,13.85,Rm}

\maketitle
%%%%%%%%%%%%%%%%%%%%%% intrudction on motivation %%%%%%%%%%%%%%%%%%%%%%%%
Supersymmetry (SUSY) postulates a symmetry between bosonic and
fermionic degrees of freedom and predicts the existence of a 
supersymmetric partner for each standard model (SM) particle.
Supersymmetric extensions of the SM provide mechanisms for solving the
hierarchy problem and offer the possibility of unification of interactions. 
An $R$-parity quantum number is defined as
$R=(-1)^{2S+L+3B}$~\cite{RParity}, where $B$, $L$ and $S$ are, respectively, the baryon
and lepton quantum numbers and the spin of the particle, such that
SM particles have $R=+1$ and their SUSY partners have $R=-1$.
$R$-parity is often assumed to be conserved, which preserves $L$ and
$B$ quantum number invariance and leaves the lightest supersymmetric particle
(LSP) stable. However, there is no fully compelling reason for the assumption of $R$-parity conservation. 
In general representations of a gauge invariant and
renormalizeable superpotential, terms of $R$-parity violation (RPV)
can be included as
\begin{eqnarray}
\label{eqn:RPVpotential} {W}_{RPV} &=& \frac{1}{2}\epsilon_{ab}
\lambda_{ijk} L_{i}^a L_{j}^b E_{k} +
\epsilon_{ab}\lambda^{\prime}_{ijk} L_{i}^a Q_{j}^b D_{k}  \nonumber \\
&+& \frac{1}{2}\epsilon_{\alpha\beta\gamma}\lambda^{\prime\prime}_{ijk}
   U_{i}^{\alpha} D_{j}^{\beta} D_{k}^{\gamma} +
\epsilon_{ab}\mu_{i} L_{i}^a H_{u}^b,
\end{eqnarray}
where $L$ and $Q$ are the lepton and quark $SU(2)$ doublet
superfields, and $E$, $U$ and $D$ denote the singlet fields. The
indices $i,j,k=1,2,3$ refer to fermion generation; $a,b=1,2$ are $SU(2)$
isospin indices; and $\alpha,\beta,\gamma=1,2,3$ are $SU(3)$ color indices. The bilinear
terms $\mu L H$ mix the lepton and the Higgs superfields, which
could yield neutrino masses and introduce a natural description of
neutrino oscillation~\cite{Rbilinear}. The trilinear terms $LLE$ and
$LQD$ represent lepton flavor violating interactions, and the $UDD$ terms lead to baryon
number violation, where interaction strengths are given, respectively, by the
dimensionless Yukawa coupling constants $\lambda$, $\lambda^{\prime}$ and
$\lambda^{\prime\prime}$.

A single slepton could be produced in hadron collisions by $LQD$
interactions and then decay into SM di-lepton final states
via $LLE$ interactions. The observation of a high-mass di-lepton
resonance would be  evidence of new physics beyond the
SM~\cite{ppem}. In this Letter, we report a direct search for resonant production 
of sneutrinos decaying into an electron and a muon in $p\overline{p}$ collisions at
$\sqrt{s}=$1.96 TeV at the Tevatron. The search is performed under the hypothesis that the
third-generation sneutrino ($\tilde{\nu}_{\tau}$) is the LSP and dominant, namely
by assuming that all couplings but $\lambda_{311}^{\prime}$ and
$\lambda_{312}=\lambda_{321}$ are zero. The final state is characterized by an electron and a muon, both of
which are well isolated and have high transverse momentum ($p_T$)
which is approximately half of the sneutrino mass.
The main background contributions are from
$Z/\gamma^*\rightarrow\tau\tau$, $WW$, $t\bar{t}$, $WZ$, and $ZZ$
processes that sequentially decay to $e\mu$ final states. High $p_T$
leptons in the signal process allows us to employ high $p_T$
thresholds to suppress the background.%The Feynman diagram is depicted in 
%FIG.~\ref{fig:feynman}.
%%%%%
%\begin{figure}
%  \includegraphics[scale=0.25]{figs/fig_feynman.eps}
%  \caption{\label{fig:feynman} The leading order Feynman diagram of subprocess 
%5           $d\overline{d}\rightarrow\tilde{\nu}_{\tau}\rightarrow \mu^{-} e^+$.}	
%\end{figure}
%A compilation of $2\sigma$ indirect bounds on couplings, under the single coupling dominance assumption with 
%a degenerate sparticle mass spectrum of $M=100$ GeV, are given in Ref.~\cite{limits} as
%\begin{equation}
%\lambda_{311}^{\prime} \le 0.12, \hskip 0.3cm \lambda_{312} \le 0.07, \hskip
%0.3cm M \equiv M_{\tilde{\nu}_{\tau}} = 100~{\text {\rm GeV}}.
%\label{eqn:lep_limits} 
%\end{equation} 

The indirect $2\sigma$ upper limit on the product of $\lambda_{311}^{\prime}\times \lambda_{312}$ from the Sindrum II experiment, reviewed by Ref.~\cite{limits}, is $2.1\times 10^{-8}$ for a degenerated sparticle mass spectrum of $M = 100$ GeV. Under the single coupling dominance assumption, where each coupling at a time is assumed to be non-zero, the indirect $2\sigma$ bounds  are as 
\begin{equation}
\lambda_{311}^{\prime} \le 0.12, \hskip 0.3cm \lambda_{312} \le 0.07, \hskip
0.3cm M \equiv M_{\tilde{\nu}_{\tau}} = 100~{\text {\rm GeV}}.
\label{eqn:lep_limits} 
\end{equation} 
A direct search for this process has been performed by  the CDF Collaboration with Tevatron Run II data~\cite{CDF}.

%%%%%%%%%%%%%%%%%%%%%% detector specifications %%%%%%%%%%%%%%%%%%%%%%%%
The D$0$ detector comprises a central tracking system in a 2 T
superconducting solenoidal magnet, a liquid-argon/uranium
calorimeter, and a muon spectrometer~\cite{d0_detector}. The
tracking system consists of a silicon microstrip tracker (SMT) and
a scintillating fiber tracker (CFT) with eight layers mounted on thin
coaxial barrels; it provides coverage for charged particles in the
pseudorapidity range $|\eta|<3$, which is defined as 
$\eta \equiv -\ln[\tan(\frac{\theta} {2})]$ where
$\theta$ is the polar angle with respect to the proton beam direction.
The calorimeter consists of a central section (CC) covering  up to
$|\eta|\approx 1.1$, and two end caps (EC) extending coverage
to $|\eta|\approx 4.2$, each housed in a separate cryostat. 
Each section consists of an inner electromagnetic (EM) compartment, 
followed by a hadronic compartment. The EM calorimeter
has four longitudinal layers and transverse segmentation of
0.1$\times$ 0.1 in $\eta-\phi$ space (where $\phi$ is the azimuthal
angle), except in the third layer, where it is 0.05$\times$ 0.05. 
The muon system resides
beyond the calorimeter and consists of a layer of tracking detectors
and scintillation trigger counters before 1.8 T iron toroidal magnets,
followed by two similar layers after the toroids. Luminosity is
measured using plastic scintillator arrays located in front of the
EC cryostats, covering $2.7<|\eta|<4.4$. The data acquisition system
consists of a three-level trigger, designed to accommodate the high
instantaneous luminosity. For final states containing an electron
with $p_T$ above 30 GeV, the trigger efficiency is close to 100\%.
  The data sample used in
this analysis was collected between April 2002 and February 2006
and corresponds to an integrated luminosity of  1.04$\pm$0.06 fb$^{-1}$.

%%%%%%%%%%%%%%%%%%%%%% event selection %%%%%%%%%%%%%%%%%%%%%%%%
%%%%%electron::
Only electrons in the CC region are considered in this analysis. The electron selection requires (i) an EM cluster with a cone of
radius  $\Delta R \equiv\sqrt{(\Delta\phi)^2+(\Delta\eta)^2} =
0.2$ in the central calorimeter, with transverse energy $E_T > 30$
GeV, where $E_T$ is defined as the cluster energy times $\sin\theta$; (ii) at least $90\%$ of the cluster energy
be deposited in the EM section of the calorimeter; (iii) the
calorimeter isolation variable ($I$) should be less than 0.15, where $I\equiv \frac
{E_{\text{tot}}(0.4) - E_{\text{EM}}(0.2)} {E_{\text{EM}}(0.2)}$, 
$E_{\text{tot}}(0.4)$ is the total energy in a cone of radius
0.4, and $E_{\text{EM}}(0.2)$ the EM energy in a cone of radius 0.2 around
the electron candidate direction; (iv) the transverse and
longitudinal shower profiles be consistent with those of electrons;
and (v) a track pointing to the EM cluster. To suppress the
misidentification of jets as electrons, an electron likelihood
discriminant based on the calorimeter variables and additional
tracking information is defined.  To ensure a high efficiency for
signal events, we impose the likelihood requirement on electron
candidates in the 30 GeV$< E_{T} <$100 GeV region, and  not the $E_T
\geq$ 100 GeV region, where the jet contamination is substantially
reduced.
 The
reconstruction efficiencies of electrons  are determined from a $Z
\rightarrow e^{+}e^{-}$ data sample to be  $(80\pm2)\%$  for $E_T<$ 100
GeV and $(86\pm2)\%$ for $E_T\geq$ 100 GeV.

%%%%%muon::
The muon candidate  is required to be separated from the
electron candidate by $\Delta R>0.2$ and from any jets by $\Delta R>0.5$,
where jets are reconstructed using an iterative seed-based cone
algorithm~\cite{jet_definition}.
In addition, we require (i) that the track $p_{T}$  be above 25 GeV; 
(ii) hits in the muon scintillation counters with time consistent with originating from the proton-antiproton collision;
(iii) at least 8 CFT hits along the track; (iv) the $E_T$ sum of the calorimeter
cells in the annulus cone of $0.1<\Delta R < 0.4$ be less than 2.5 GeV, and the transverse momentum sum
of all tracks besides the muon track within a cone of radius $\Delta R = 0.5$ be less than 2.5 GeV. The
reconstruction efficiency of muons determined from a $Z \rightarrow
\mu^{+}\mu^{-}$ data sample is  $(81\pm2)\%$.

%%%%%veto::
%Events with electron and muon candidate are demanded to pass
%respective multi-lepton, jet and missing transverse energy veto
%cuts as:(1) to suppress WZ and ZZ backgrounds, events are rejected
%if an additional third lepton is found as electron $E_{T}>$8 GeV
%without likelihood discriminant or muon $p_{T}>$5 GeV; (2) to
%suppress $t\bar{t}$ background, events are rejected if a jet
%defined using an iterative seed-based cone
%algorithm\cite{jet_definition} is found in $|\eta|<$2.5 region
%with $p_{T}>$30 GeV; (3) any events with missing transverse energy
%$\slash{E}_{T}>$15 GeV and away from the muon candidate as
%0.6$<\Delta\phi({\slash E}_T,\mu)<$2.5 will be removed, where
%$\slash{E}_{T}$ is calculated as the negative vector sum of energy
%deposits in the calorimeter cells, taking into account energy
%corrections for reconstructed electrons, muons and jets.
To suppress $WZ$ and $ZZ$ background, events having two muon
candidates with $p_T>$ 5 GeV or two electron candidates with $p_T>$ 8
GeV are rejected. In order to suppress the $t\bar{t}$ background, events with missing transverse energy $\slashed{E}_T > 15$ GeV that is not aligned or antialigned in azimuth with the muon ($0.6<\Delta\phi(\slashed{E}_T,\mu) <2.5$ rad), as well as events with at least one jet with $p_T > 30$ GeV and $|\eta| < 2.5$ are rejected.  

% or  missing transverse energy $\slashed{E}_T > 15$ GeV are
%rejected. For the $\slashed{E}_T$ veto, events with $\slashed{E}_T$ close to the muon direction  in the $x-y$ plane($\Delta\phi(\slashed{E}_T,\mu) < 0.6$ rad or $\Delta\phi(\slashed{E}_T,\mu) >2.5$ rad, where $\Delta\phi(\slashed{E}_T,\mu)$ denotes the relative azumuthal angle between the muon and the $\slashed{E}_T$) are exempted from the veto. Events with at least one jet with $p_T$ above 30 GeV are rejected in
%order to suppress the $t\bar{t}$ background.

%%%%%%%%%%%%%%%%%%%%%%% event generator %%%%%%%%%%%%%%%%%%%%%%%%%%
The partonic signal events are generated using the
 {\sc {comphep}} program~\cite{CompHEP} and CTEQ6L~\cite{CTEQ} parton distribution functions (PDF).
The cross section of the process depends on sneutrino mass $M$ and 
the $LQD$ and $LLE$ coupling constants as \cite{ppem}
\begin{equation}
  \hat{\sigma}_{e\mu} \propto
   (\lambda^{\prime}_ {311})^{2} \times (\lambda_{312})^{2} \cdot
   \frac {1} {|\hat{s}- M^2+i\Gamma M|^2},
  \label{eqn:sigmaRPV}
\end{equation}
where $\Gamma$, the total width of the LSP sneutrino, includes all decay modes ($d\overline{d}$ and $e\mu$), and also depends on the $LQD$ and $LLE$ couplings  as
\begin{equation}
  \Gamma = [ 3 \cdot (\lambda^{\prime}_{311})^2 + 2 \cdot (\lambda_{312})^2 ] \cdot \frac{M}{16 \pi}.
  \label{eqn:GammaRPV}
\end{equation}
A mass-dependent $K$-factor is applied to include
next-to-leading order QCD corrections~\cite{ppemQCD}. 
The partonic signal events are processed through {\sc {pythia}}~\cite{PYTHIA}
to include parton showering, hadronization and particle decays.
The influence of the PDF uncertainty on the cross section times acceptance is 6.2\% -- 8.6\% depending on the sneutrino mass, estimated from the CTEQ6M error functions. The cross section uncertainty from the choice of renormalization scale and factorization scale is about 4\%. 
Standard model background processes are generated with {\sc {pythia}} and CTEQ6L1. The
contribution of Drell-Yan $Z/\gamma^*$ processes is normalized using
the NNLO cross section~\cite{Drell-Yan-X}. The contributions of
$WW$, $WZ$ and $t\bar{t}$ processes are normalized with NLO cross
sections~\cite{diboson-X, ttbar-X}.
%The next-to-leading-order (NLO) cross section of $t\bar{t}$ production
%at Tevatron is 6.1pb~\cite{ttbar-X}, and NLO cross sections of WW,
%and WZ~\cite{diboson-X} are 12.1pb and 3.7pb respectively.
%%%
%Besides above SM contribtions, reducible backgrounds
%such as W+jets with one of the jets fakes an electron candidates, W+$\gamma$ with photon conversion and multi-jet events proved to be
%negligible in data, due to our tight event selection strategy,
%%%
All signal and background events are processed with a detailed {\sc
{geant}}-based D$0$ detector simulation~\cite{D0-simulation}, and
are corrected for trigger effects and the differences in the
reconstruction efficiencies compared to those in data.
The background from misidentification of photons or jets as leptons, such as $W\gamma$ and $W+$jet
and QCD di-jet events, is estimated from data and is found to be negligible given our stringent event selection criteria.

\begin{figure}
  \includegraphics[width=8.6cm]{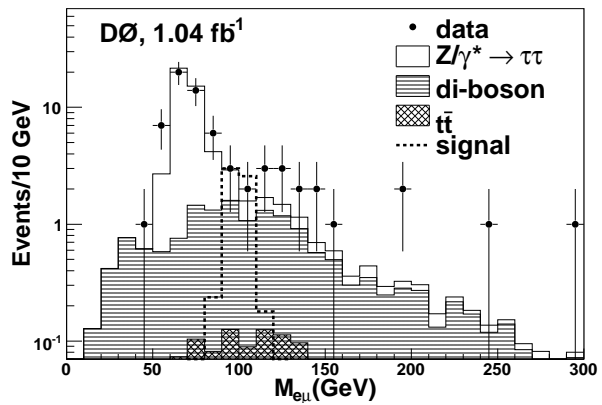}
  % Here is how to import EPS art
  \caption{\label{fig:Memu}  Invariant mass of the electron-muon system.
    The di-boson  contribution includes the
    $WW$ and $WZ$ processes. The dashed  line indicates the signal Monte Carlo
    simulation of sneutrino with mass of 100 GeV and $\sigma\times$BR of 0.057 pb. }	
\end{figure}

The number of selected events in data and the estimated background contributions
are summarized in Table~\ref{tab:sum}. The $ZZ$ contribution is found to be negligible
after the event selection and is not listed.
%%%%%
\begin{table}
  \caption{\label{tab:sum} The numbers of selected events in data and 
    different estimated background contributions.}
  % \begin{ruledtabular}
  \begin{tabular}{cc}
    \hline \hline
  Process & Events\\
    \hline
    $Z/\gamma^*\rightarrow\tau\tau$ & $42.9 \pm 4.2$\\
    $WW$  & $13.7 \pm 1.5$\\
    $t\bar{t}$ & $1.4 \pm 0.3$\\
    $WZ$ & $1.2 \pm 0.2$\\
    \hline
    Total background  & $59.2 \pm 5.3$\\
    \hline 
   Data & 68\\
   \hline
   \hline
  \end{tabular}
  % \end{ruledtabular}
\end{table}
There are 68 candidates found in the data. The expectation from the
SM processes is $59.2 \pm 5.3$ events, where the uncertainty includes the
statistical uncertainty and uncertainties from the integrated
luminosity (6\%), reconstruction and trigger efficiencies (3.1\%),
and background cross sections ($Z/\gamma^*$ (3.5\%), $t\bar{t}$ (14.7\%), and 
di-boson production (5.6--6.6\%)). The kinematic
variables of the final state are well described by the sum of the SM
background contributions.
%There is a good c on
The distribution of the electron and muon invariant mass ($M_{e\mu}$) is shown in Fig.~\ref{fig:Memu}.

\begin{figure}
  \includegraphics[height=5.4cm]{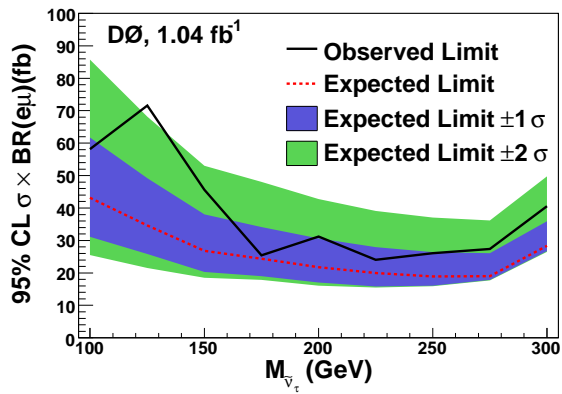}
  \caption{\label{fig:xsection} The observed and expected upper limits on 
$\sigma\times BR$ at 95\% CL for the process $p\overline{p}\rightarrow\tilde{\nu}_{\tau} + X\rightarrow e\mu + X$ as a function of the sneutrino mass, assuming that the sneutrino total width is much narrower than our detector resolution (color online).}
\end{figure}

\begin{figure}[floatfix]
  \includegraphics[height=5.2cm]{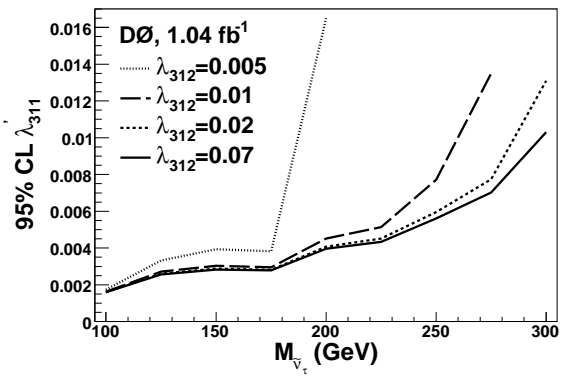}
  \caption{\label{fig:couplings} The observed upper limits on
	    $\lambda^{\prime}_{311}$  at 95\% CL for four fixed values of $\lambda_{312}$ as a function of the sneutrino mass.  
}
\end{figure}

 Using the $M_{e\mu}$ distributions, we calculate an upper limit on 
$\sigma\times BR$ for the process
$p\overline{p}\rightarrow\tilde{\nu}_{\tau} + X\rightarrow e\mu + X $ with a modified frequentist ($CL_s$) method~\cite{lhood-ratio}, under the assumption that the total width is much narrower than the detector resolution.  
The upper limits as a function of the sneutrino mass are shown in  Fig.~\ref{fig:xsection}.
 We fix one of the coupling constants  and set the upper limit on the other 
for different sneutrino masses. Shown in Fig.~\ref{fig:couplings} are the observed upper limits on $\lambda^{\prime}_{311}$ for four assumed values of $\lambda_{312}$. For a sneutrino with mass of 100 GeV, $\lambda^{\prime}_{311}>1.6\times10^{-3}$ is excluded at 95\% CL when $\lambda_{312}=0.01.$ 
%MC expectations, as depicted FIG.~\ref{fig:Memu}
%%%%%

%%%%%
%Under assumption that the number of observed events is consistent with predicted background, we get
%95\% C.L. exclusion limits of $\sigma\times$BR for a generic scalar boson decaying into $e\mu$ final state,
%using a modified frequentist ($CL_s$) method~\cite{lhood-ratio} on the invariant mass distribution. 

%%%%%
%\begin{table}
%  \caption{\label{tab:CLsb} The observed and expected upper limits on RPV couplings
%    for differen sneutrino masses at 95\% C.L. }
%  \begin{ruledtabular}
%    \begin{tabular}{ccc}
%      $M_{\tilde {\nu}_{\tau}}$ & $\lambda^{,}_{311} \times
%      \lambda_{312}$
%      & $\lambda^{,}_{311} \times \lambda_{312}$\\
%      (GeV) & (observed limits) & (expected limits)\\
%      \hline
%      100 & $3.01 \times 10^{-3}$ & $2.57 \times 10^{-3}$\\
%      150 & $3.40 \times 10^{-3}$ & $3.00 \times 10^{-3}$\\
%      200 & $4.40 \times 10^{-3}$ & $3.96 \times 10^{-3}$\\
%      250 & $5.26 \times 10^{-3}$ & $4.98 \times 10^{-3}$\\
%      300 & $6.77 \times 10^{-3}$ & $6.81 \times 10^{-3}$\\
%    \end{tabular}
%\end{ruledtabular}
%\end{table}
%%%%%

%Accordingly, the exclusion region in the RPV coupling constants versus sneutrino mass is depicted in 
%Fig. \ref{fig:couplings}. The plot demonstrates that sensitivity on parameters is greatly improved, 
%for example, $\lambda^{,}_{311}>1.6\times10^{-3}$ is excluded at 95\% C.L. for sneutrino with mass of 
%100 GeV and $\lambda_{312}$ about 0.01.

In summary, we have studied the production of high $p_T$
electron-muon pair final states with about 1 fb$^{-1}$ of D$0$
 data. We select 68 events, while the SM expectation is
59.2$\pm$5.3 events. The distributions of kinematic variables are in
good agreement with the SM predictions. We set limits on the
parameters of a particular supersymmetric model which predicts an
enhancement of the high $p_T$ electron-muon final state via
$R$-parity violating production and decay of sneutrino particles. These are the most
stringent direct limits to date.
%The constraints significantly improve on previous results~\cite{limits} 
%for a sneutrino with mass of 100 GeV.

% acknowledgement_paragraph_r2.tex                                 2/28/07
% acknowledgement_paragraph_r2.tex                                 10/09/07
%
We thank the staffs at Fermilab and collaborating institutions, 
and acknowledge support from the 
DOE and NSF (USA);
CEA and CNRS/IN2P3 (France);
FASI, Rosatom and RFBR (Russia);
CAPES, CNPq, FAPERJ, FAPESP and FUNDUNESP (Brazil);
DAE and DST (India);
Colciencias (Colombia);
CONACyT (Mexico);
KRF and KOSEF (Korea);
CONICET and UBACyT (Argentina);
FOM (The Netherlands);
Science and Technology Facilities Council (United Kingdom);
MSMT and GACR (Czech Republic);
CRC Program, CFI, NSERC and WestGrid Project (Canada);
BMBF and DFG (Germany);
SFI (Ireland);
The Swedish Research Council (Sweden);
CAS and CNSF (China);
Alexander von Humboldt Foundation;
and the Marie Curie Program.
%%%% remove Marie Curie at August 07 update
%

%%%%%%%%%%%%%%%%%%%%%%%%%%%%%%%%%%%%%%%%%%%%%%%%%%%%%%%%%%%%%%%%%%%%%%

\end{document}